\documentclass[twocolumn,preprintnumbers,amsmath,amssymb,superscriptaddress]{revtex4-1}
\usepackage{graphicx}
\usepackage{dcolumn}
\usepackage{soul}
\usepackage[usenames,dvipsnames]{xcolor}
\usepackage{mathptmx}
\usepackage{amssymb}
\usepackage{array}
\usepackage{amsmath}
\usepackage{bm}
\usepackage{setspace}
\usepackage{hyperref}

\begin{document}

\title{Magnetically Assisted Trapping of Passive Colloids by Active Dipolar Chains}

\author{A. Compagnie}
\address{Institut f\"ur Theoretische Physik: Weiche Materie, Heinrich-Heine-Universit\"at, D-40225, D\"usseldorf, Germany.}
\address{GRASP, Institute of Physics B5a, University of Li\`ege, B4000 Li\`ege, Belgium.}
\author{N. Vandewalle}
\address{GRASP, Institute of Physics B5a, University of Li\`ege, B4000 Li\`ege, Belgium.}
\author{E. Opsomer}
\address{GRASP, Institute of Physics B5a, University of Li\`ege, B4000 Li\`ege, Belgium.}


\begin{abstract}

We investigate a trapping mechanism for passive Brownian particles based on mixtures with self-propelled dipolar colloids. Active dipoles, whose magnetic moment is oriented perpendicularly to their propulsion direction, spontaneously form dynamic chains that collapse into clusters through dipole–dipole interactions. These transient structures efficiently capture nearby passive particles, forming dense phases at relatively low global densities. Using Brownian dynamics simulations, we analyze how the capture efficiency depends on the Péclet number ($Pe$) and dipolar interaction strength ($\lambda$). We demonstrate that an external magnetic field, applied briefly to align the active dipoles, significantly enhances trapping efficiency, with capture rates exceeding 50\% under optimal conditions. Our results reveal a nontrivial competition between activity and dipolar forces, governed by the ratio $\lambda/Pe$, and offer insights into designing self-organized trapping strategies for passive colloids. 

\end{abstract}
\maketitle


\section{Introduction}

Active systems, characterized by their ability to extract energy from their environment and convert it into motion \cite{elgeti_physics_2015}, can self-organize into configurations that are unattainable in equilibrium conditions \cite{vicsek_collective_2012}. Originally observed in biological contexts \cite{dombrowski_self-concentration_2004,sokolov_concentration_2007,cavagna_bird_2014}, such behavior has since also been extensively investigated using artificial self-propelled particles, both experimentally \cite{kudrolli_swarming_2008,deseigne_collective_2010,walther_janus_2013} and through numerical and theoretical approaches \cite{bechinger_active_2016,mallory_active_2018,solon_active_2015,martin_statistical_2021}. One of the most striking collective phenomena emerging from activity is Motility-Induced Phase Separation (MIPS) \cite{cates_motility-induced_2015}. While MIPS can arise solely from activity and steric interactions among spherical particles \cite{fily_athermal_2012,redner_structure_2013,bialke_microscopic_2013}, the collective dynamics of the particles can be altered by considering additional features such as non-circular particles \cite{suma_motility-induced_2014,shi_self-propelled_2018}, phoretic interactions \cite{pohl_dynamic_2014,liebchen_phoretic_2017,stark_artificial_2018}, local alignment \cite{vicsek_novel_1995,barre_motility-induced_2015,martin-gomez_collective_2018}, or self-alignment mechanisms \cite{baconnier_self-aligning_2025,musacchio_self-alignment_2025}.

Dipolar interactions between active colloids are of particular interest, due to the specific collective properties they can exhibit \cite{klapp_collective_2016}. They may come from electrostatic imbalance, induced dipoles or permanent dipoles, collectively leading to a variety of possible configurations \cite{kaiser_active_2015,yan_reconfiguring_2016,liao_dynamical_2020,telezki_simulations_2020}, such as chains, rings or clusters. This can also be modified by tuning these interactions, for example by shifting the position of the dipole \cite{yener_self-assembly_2016}, by having induced dipoles perpendicular to the self-propulsion \cite{chao_traveling_2025}. Another advantage of magnetic dipoles, in the optics of external control of active systems, is their response to external magnetic fields. This has been shown to have significant impact on the organization and the dynamics of the system \cite{telezki_patterns_2025,parage_modulation_2025}.

Complex collective behaviors also stem from interaction between population of particles with varying characteristics, which will form mixtures \cite{kolb_active_2020}. This is already observed when considering simple differences, as in active-passive mixtures where MIPS occur with passive particles trapped in dense phases \cite{stenhammar_activity-induced_2015,wittkowski_nonequilibrium_2017,dolai_phase_2018,rogel_rodriguez_phase_2020,gokhale_dynamic_2022}, but other effects may arise, such as the demixing of chiral active particles due to different chiral orientations \cite{mijalkov_sorting_2013,ai_spontaneous_2023}. By varying multiple characteristics, even more different collective phenomena may occur \cite{mccandlish_spontaneous_2012,klapp_collective_2016,agudo-canalejo_active_2019,sturmer_chemotaxis_2019,maloney_clustering_2020,maloney_clustering_2020-1}.

This line of research is not only of fundamental interest but also relevant for practical applications. Inert contaminants, such as microplastics suspended in aqueous environments, can be effectively modeled as passive particles in a viscous medium. Developing self-organized strategies to trap such particles using active agents could open new avenues in solvent purification and environmental remediation \cite{gao_seawater-driven_2013,jurado-sanchez_self-propelled_2015}. However, activity alone is not always sufficient to achieve efficient trapping of passive components, particularly at the low concentrations required for spontaneous MIPS formation in mixed systems \cite{wittkowski_nonequilibrium_2017}.

In this work, we investigate how dipolar interactions among active particles can assist in the formation of functional structures capable of efficiently capturing passive particles, even at low densities. By characterizing the formation mechanisms, internal structure, and dynamics of these active–passive aggregates, we aim to identify robust and spontaneous trapping strategies.

The remainder of the paper is organized as follows. Section II introduces the model and key parameters used in the simulations. In Section III, we present the trapping results under various conditions, with the aim of optimizing capture efficiency. Section IV offers a detailed analysis of how activity and dipole–dipole interactions shape the observed dynamics. Finally, Section V summarizes the main findings and discusses potential future directions.

\section{Model}

Our system consists in $N$ colloids that we model as disks of diameter $\sigma$ that are placed randomly in a fluid bath of viscosity $\eta$. These colloids form two separate populations because of their differing properties; $N_a$ colloids are active Brownian particles with a central magnetic dipole, and $N_p$ colloids are passive Brownian particles. Due to their characteristics, active disks are called active dipoles. Providing that inertial effects can be neglected, the system dynamics can be described by the following stochastic differential equations,
\begin{align}
     \dot{\mathbf{r}}_i &= v_i \mathbf{\hat{n}}_i + \sqrt{2D_T} \ \bm{\xi}_T - \frac{\nabla_i U(\mathbf{r}_i,\theta_i)}{3\pi\eta\sigma},\\
     \dot{\theta_i} &= \sqrt{2D_R} \ \xi_R - \frac{\partial_{\theta_i} U(\mathbf{r}_i,\theta_i)}{\pi\eta\sigma^3} + \frac{M_{B, i}}{\pi\eta\sigma^3},
\end{align}
where $\bm{r}_i$ is the disk's position, $\theta_i$ its orientation with $\mathbf{\hat{n}}_i = (\cos{\theta_i}, \sin{\theta_i})^T$, and $v_i$ its self-propulsion velocity with $v_i = v_0$ for active dipoles and $v_i = 0$ for passive colloids. Parameters $D_T$ and $D_R$ denote the translational and rotational diffusion coefficients respectively, and $\bm{\xi}_T$ and $\xi_R$ correspond to Gaussian white noises with unit variance and zero mean. The interaction potential $U(\mathbf{r}_i,\theta_i)$ is given by 
\begin{align}
    U(\mathbf{r}_i,\theta_i) =  \sum_{j\neq i}^N\left(u_{WCA}(|\mathbf{r}_{ij}|) + u_{dd}(\mathbf{r}_{ij}, \theta_i, \theta_j)\right),
\end{align}
with $\mathbf{r}_{ij}=\mathbf{r}_i - \mathbf{r}_j$. The first term corresponds to the short-ranged and repulsive Weeks-Chandler-Andersen (WCA) potential, 
\begin{align}
    u_{WCA}(r_{ij}) = 
    \begin{cases}
        4\epsilon \left( ( \frac{\sigma}{r_{ij}} )^{12} + ( \frac{\sigma}{r_{ij}} )^6 \right) + \epsilon & \text{if } r_{ij} < 2^{1/6} \sigma \\
        0 & \text{otherwise}
    \end{cases}
\end{align}
with $\epsilon$ the energy scale of the contacts. The second term corresponds to the magnetic dipole-dipole interaction,
\begin{align}
    u_{dd}(\mathbf{r}_{ij}, \theta_i, \theta_j) &= \frac{\mu_0}{4\pi} \frac{\bm{\mu}_i \cdot \bm{\mu}_j - 3 (\bm{\mu}_i \cdot \mathbf{\hat{r}}_{ij}) (\bm{\mu}_j \cdot \mathbf{\hat{r}}_{ij})}{r_{ij}^3}\\
    \bm{\mu}_i &= \mu_i (\cos(\theta_i+\phi_i), \sin(\theta_i+\phi_i))^T
\end{align}
where $\mu_0$ is the vacuum permeability, $\mu_i$ the magnetic dipole moment with $\mu_i = \mu$ for active dipoles and $\mu_i = 0$ for passive colloids, and $\phi_i$ is the angle between $\bm{\mu}_i$ and $\mathbf{\hat{n}}_i$ which will be fixed to 0 or to $\pi$. An external magnetic field $\mathbf{B}$ may also be applied to the entire system  The corresponding magnetic torque exerted on the dipoles (and thus the disks) is given by
\begin{align}
    M_{B, i} = - \bm{\mu}_i \cdot \mathbf{B}
\end{align}

The global dynamics of the system will be mainly determined by the activity and the dipole-dipole interactions. Therefore, the Péclet number Pe and the magnetic dipolar adimensional number $\lambda$ are relevant to estimate the behavior of the system. 
They are defined as
\begin{align}
    \text{Pe} &= \frac{2v_0}{\sigma D_r}\\
    \lambda &= \frac{\mu_0}{4\pi}\frac{\mu^2}{\pi \sigma^3 k_B T}
\end{align}
with the different parameters chosen from the active population of colloids.

\section{Results}

\subsection{Numerical simulations}

Brownian dynamics simulations, using the Euler-Maruyama method with random Gaussian white noise generated by the PCG algorithm \cite{oneill_pcg_2014}, were conducted to investigate this model. A total of $N_a=1000$ active dipoles and $N_p=1000$ passive colloids are randomly placed in a 2D squared box before the dynamics are evaluated with periodic boundary conditions. This represents concentrations of 0.05 for each population in the system. The dipoles have their orientation shifted by $\pi/2$ compared to the direction of the self-propulsion.

The general protocol for each simulation is the same: the colloids are randomly placed in the space, a magnetic field $\mathbf{B} = B_0 \mathbf{e}_x$ may be applied for a duration of $t_B$ before being turned off, and the colloids are left to move freely during $20\tau_R$. Only three parameters may vary across the simulations: the self-propulsion velocity $v$ and the magnetic dipole moment amplitude $\mu$ of the active dipoles, and the duration $t_B$ during which the magnetic field is turned on.

In order to reduce the computational time required for these simulations, each colloid is linked to a square cell in the 2D space to only consider contacts and dipolar interactions with other particles already determined to be potential neighbors. While having no impact on the results for the calculation of contacts, since the cut-off for WCA potential is at $2^{1/6}\sigma$, this cut-off has an impact on the dynamics resulting from dipole-dipole interactions. A compromise between accuracy and efficiency was chosen to set the size of the cells to always consider dipole-dipole interactions up to 1/100th of the maximum interaction between touching colloids with aligned dipoles.

\subsection{Structures of active dipoles}

\begin{figure}[t]
    \centering
    \includegraphics[width=\linewidth]{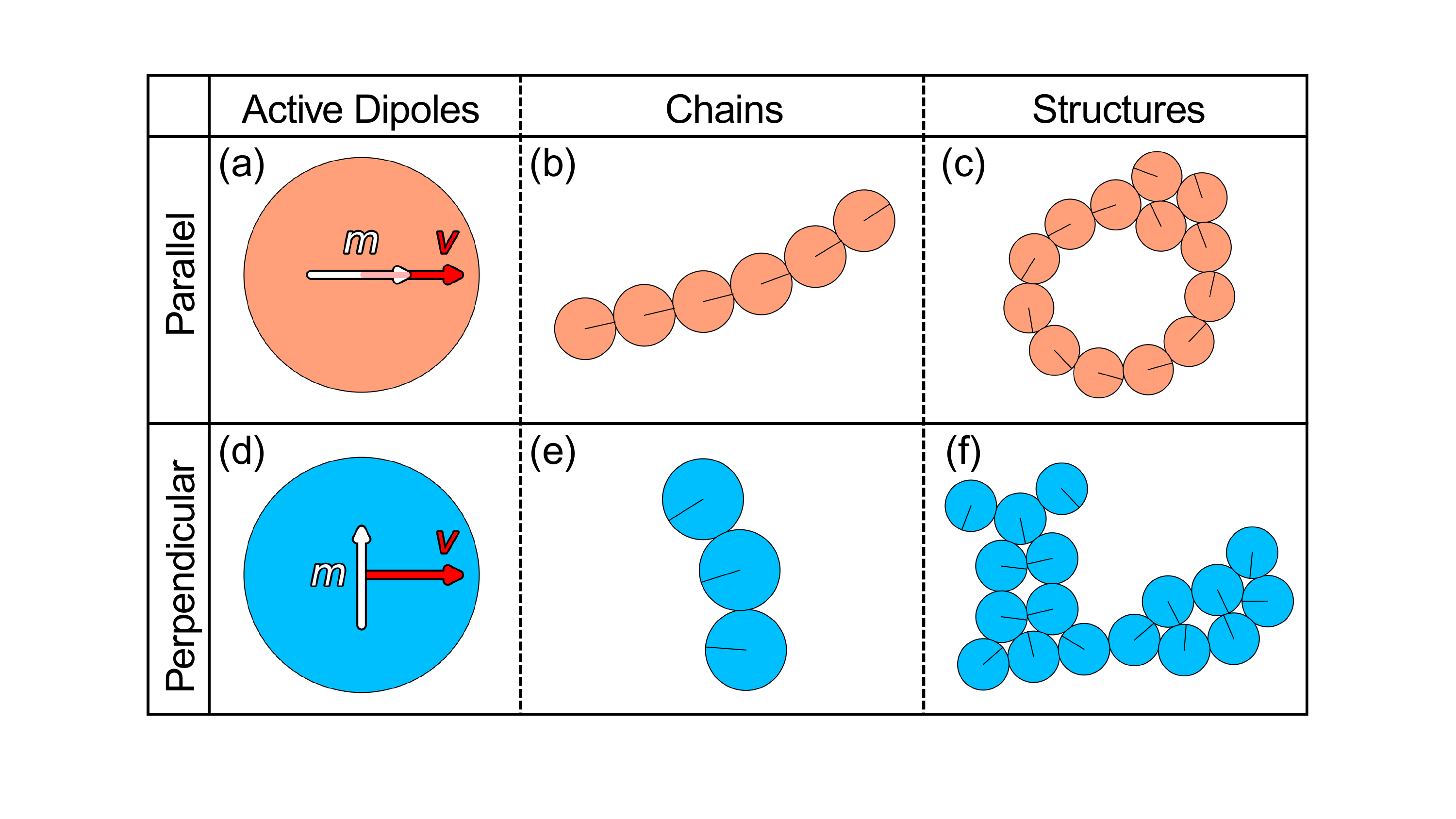}
    \caption{Representation of (a) parallel and (d) perpendicular active dipoles. The orientation of each particle, and therefore the direction of the self-propulsion, is represented by the internal black line of each disk. Parallel active dipoles form (b) chains moving along their direction, (c) which may collapse in rotating spirals. Perpendicular active dipoles form (e) small chains moving perpendicular to their length, (f) which may collapse in clusters.}
    \label{fig:activedipolestypes}
\end{figure}

Active dipoles have a tendency to form chains of particles wandering through the 2D space. If the dipolar interactions are strong enough compared to the Brownian noise, these structures are stable until a perturbation makes the chain collapse on themselves. This perturbation is usually caused by the collision with active dipoles, especially with higher concentrations of particles.

These phenomena happen whether the dipoles are oriented parallel or perpendicular to the propulsion. However, the way active dipoles are organized inside these structures is different, which leads to additional ways to cause the collapse of the chains, as shown in figure \ref{fig:activedipolestypes}.

On one hand, parallel active dipoles form chains that move along the direction of the dipoles \cite{liao_dynamical_2020,telezki_simulations_2020}. When perturbed, these chains may form spinning spirals that can become chains again over time. Particles are rarely completely stopped in the low concentrations systems, forming chains that evolve continuously in size and shape.

On the other hand, perpendicular dipoles may form chains moving perpendicularly to their length. However, these structures are less stable compared to chains of active particles with parallel dipoles. They will quickly form clusters with other chains and particles, which are extremely stable comparatively. With time, such structures may grow bigger and more passive, because the propulsion of the colloids tend cancel out each other.

\subsection{Trapping with active dipoles}

Creating dense phases of passive colloids is possible through MIPS in mixtures of active and passive colloids. Yet this phenomenon is only observed at relatively high densities \cite{stenhammar_activity-induced_2015,wittkowski_nonequilibrium_2017}. Using perpendicular active dipoles instead of non-magnetic active colloids may be a solution to form these dense phases of passive particles due to the collective behavior of these specific active dipoles.

\begin{figure}
    \includegraphics[width=0.95\linewidth]{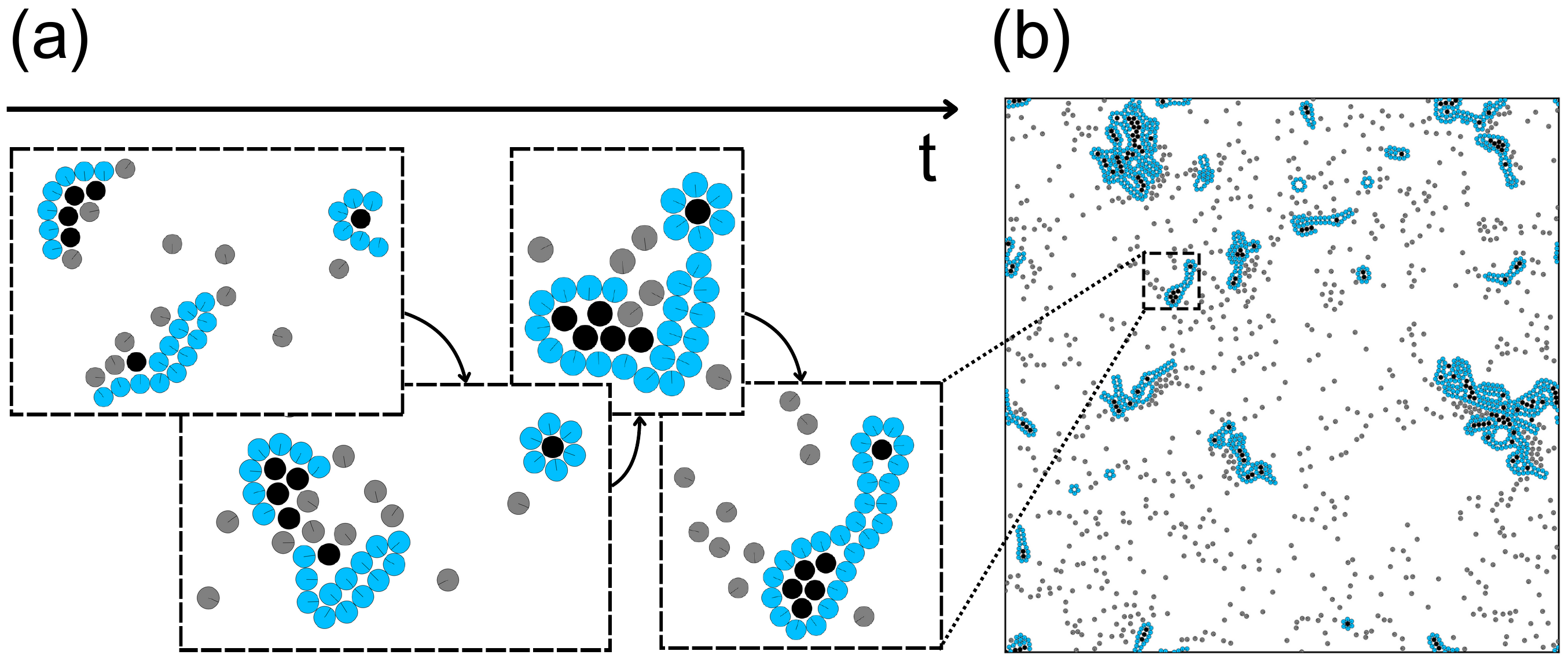}
    \caption{(a) Various snapshots illustrating the evolution of structures formed by the active dipoles (colored in blue) over time, with passive colloids colored in gray when free and black when captured. (b) Final snapshot of the simulation for which the capture rate is 0.148.}
    \label{fig:snapshots_nofield}
\end{figure}

Indeed, perpendicular active dipoles may collapse on each other due to collision with each other, but also by colliding with passive particles. Due to their ability to form small chains moving perpendicularly to their length, active dipoles can collect passive particles and trap them once they collide with other actives particles or chains. This leads to some passive particles being trapped in clusters formed by active dipoles, illustrated in figure \ref{fig:snapshots_nofield}.

\begin{figure}
    \centering
    \includegraphics[width=0.9\linewidth]{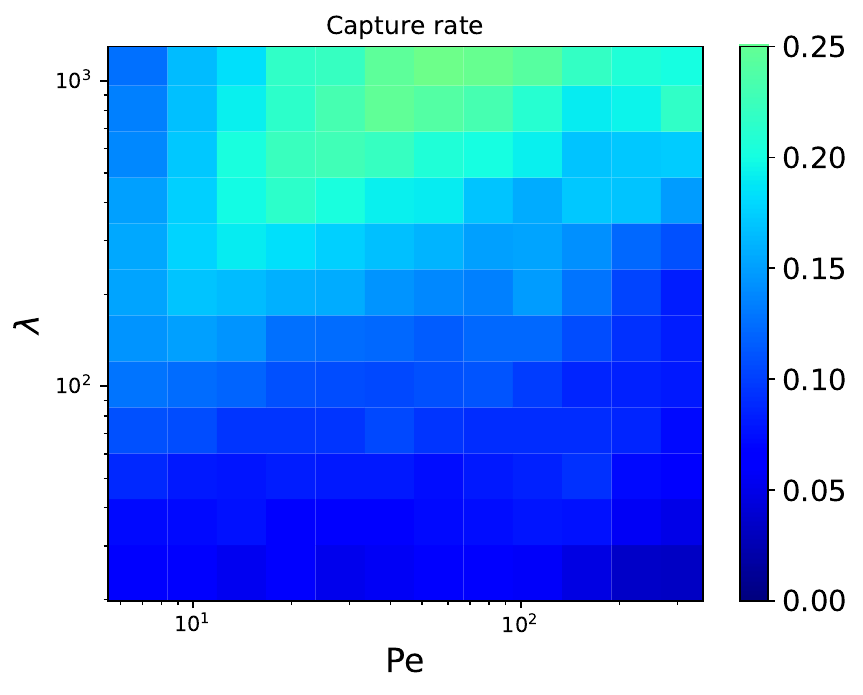}
    \caption{Capture rate histogram depending on adimensional numbers Pe and $\lambda$. Each data point is averaged over 20 simulations.}
    \label{fig:phasediagram_nofield}
\end{figure}

The efficiency of such a protocol is presented in figure \ref{fig:phasediagram_nofield}. Passive particles are considered captured in simulations by imposing a constant velocity to active particles and determining which passive particles follow in each direction. In the range of values studied for the parameters Pe and $\lambda$, there seems to be a peak efficiency of 0.25 by having the strongest dipolar interactions possible, but a specific balance for the activity is required to maximize the capture. Nonetheless, a vast majority of the passive colloids are not stuck in dense phases, even in the best case. This is because the chains are not stable enough to form chains that are stable during enough time to interact with passive colloids before collapsing in inert clusters, which are unable to capture additional passive particles.


\subsection{Impact of magnetic field}

\begin{figure}
    \centering
    \includegraphics[width=0.9\linewidth]{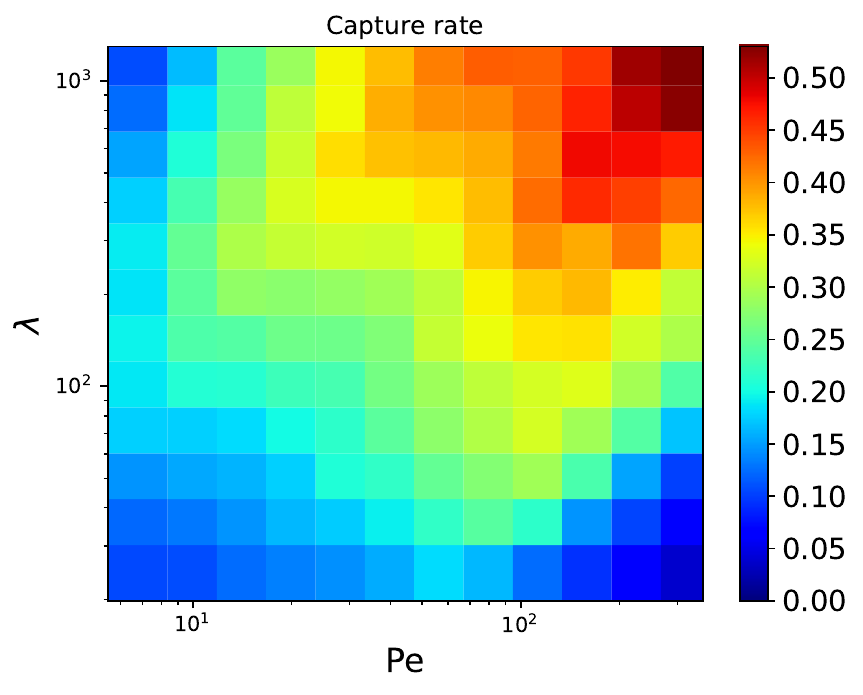}
    \caption{Capture rate histogram depending on adimensional numbers Pe and $\lambda$, after a uniform magnetic field is turned on for a duration $t_B \approx \tau_R$. Each data point is averaged over 20 simulations.}
    \label{fig:phasediagram_field}
\end{figure}

\begin{figure*}[t]
    \centering
    \includegraphics[width=0.95\linewidth]{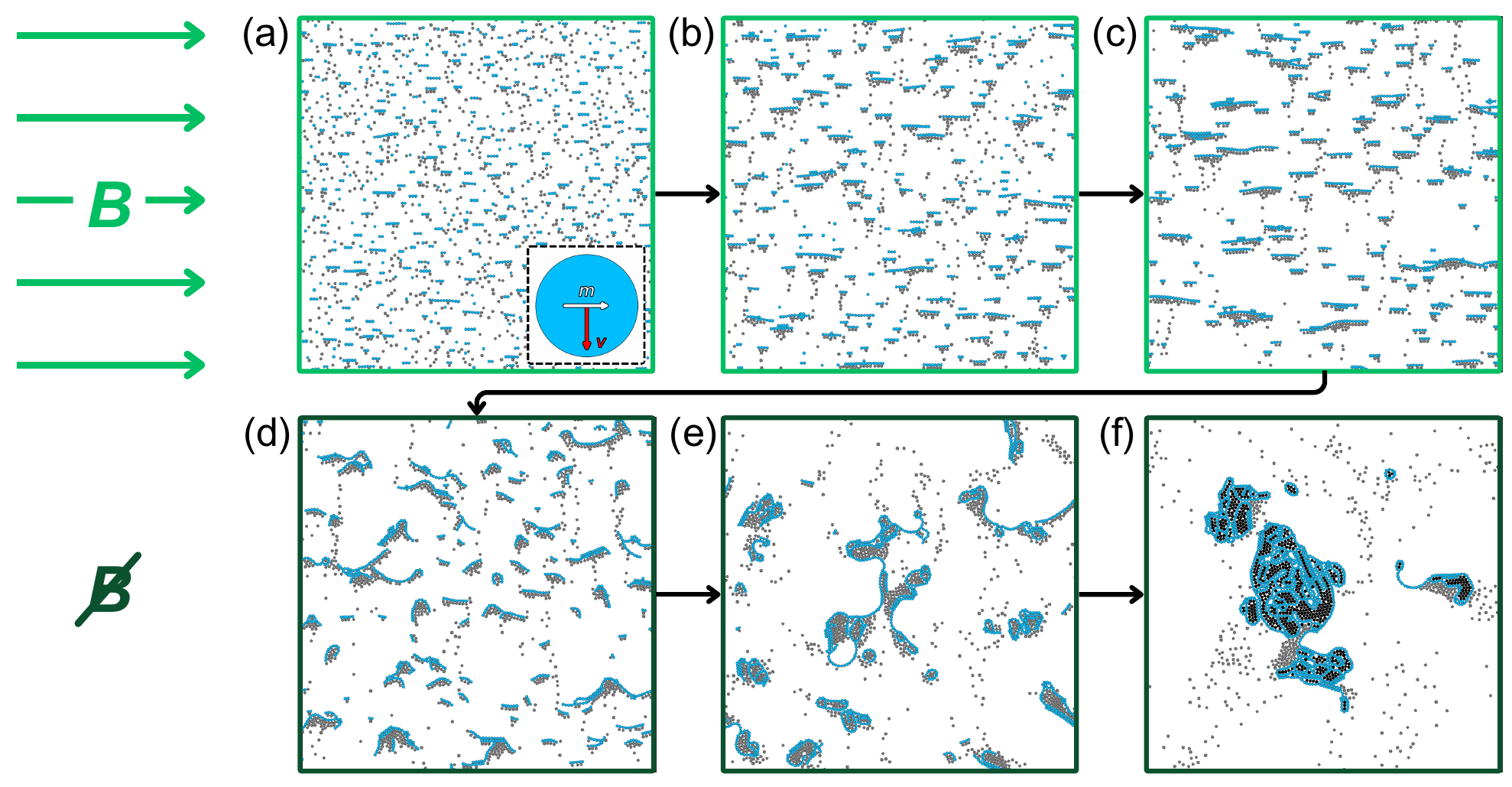}
    \caption{Representation of the capture of passive particles by dipolar active particles with a magnetic field. (a,b,c) Chains of active dipoles growing with time passing, sweeping passive particles. (a) Active dipoles have their dipole (white) aligning with the magnetic field, forcing their orientation to self-propel downwards (red). (d,e,f) Once the magnetic field is turned off, chains bend and fuse while capturing passive colloids in bigger structures. (f) Final snapshot of a simulation with captured passive colloids in black, leading to a capture rate of 0.571.}
    \label{fig:snapshots_field}
\end{figure*}

One way to improve the efficiency of this protocol is to create more stable chains. This can be done by applying an external magnetic field. Indeed, when this field is applied, an additional torque will affect each active dipoles and they will collectively have a tendency to orient themselves in one direction. This leads to a great increase in capture once the magnetic field is turned off, as shown in figure \ref{fig:phasediagram_field}. Comparing the results to the same simulations without magnetic field \ref{fig:phasediagram_nofield}, it becomes obvious that such protocol greatly improves capture in all cases, but especially for system with high values of Pe and $\lambda$, with capture rates reaching more than 0.5 on average.

Analyzing the evolution of the structures over time explains the reason different regions of high and low capture are observed. This is depicted in figure \ref{fig:snapshots_field}. During the first phase, the magnetic field is turned on, leading to all active dipoles to orient themselves and propel in the same direction (a). They will also be attracted to other active dipoles next to them, since they have similar orientations, forming small chains. With time (b,c), these chains grow bigger by fusing with others and start to sweep passive particles along the way. When the magnetic field is turned off (d), the active dipoles do not have the external orientational constraint, meaning that they will start to change shape and rotate due to noise and imbalance of self-propulsion across the structure. The chains will become more round and start to collide with each other, circling around passive colloids efficiently (e). This continuously happens until the structures get big enough so that they are mostly inert, and reach a final capture rate (f).

By increasing $\lambda$, chains and structure grow bigger and are more stable, which is a positive contribution to the capture. And by increasing Pe, the active dipoles interact faster with the passive colloids, which also increases the number of passive colloids captured in the clusters in the same time frame.

This protocol overlooks the possibility that the activity leads to different scaling in time. Indeed, by scaling $t_B$ with $\tau_R$, it is kept constant for all couples $(Pe,\lambda)$. However, the time required for chains to form and for passive colloids to be swept away directly depends on the self-propulsion $v_0$. This means that in this case, the magnetic field is not turned off after a sufficient time for chains to form in low activity systems. The easiest way to make the results more comparable for different activities is to consider that the phenomena are directly proportional to the activity. Scaling the duration of the magnetic field as $t_B \propto \sigma/v_0$ should therefore reduce the varying contribution of $t_B$ to the final results.

\begin{figure}
    \centering
    \includegraphics[width=0.9\linewidth]{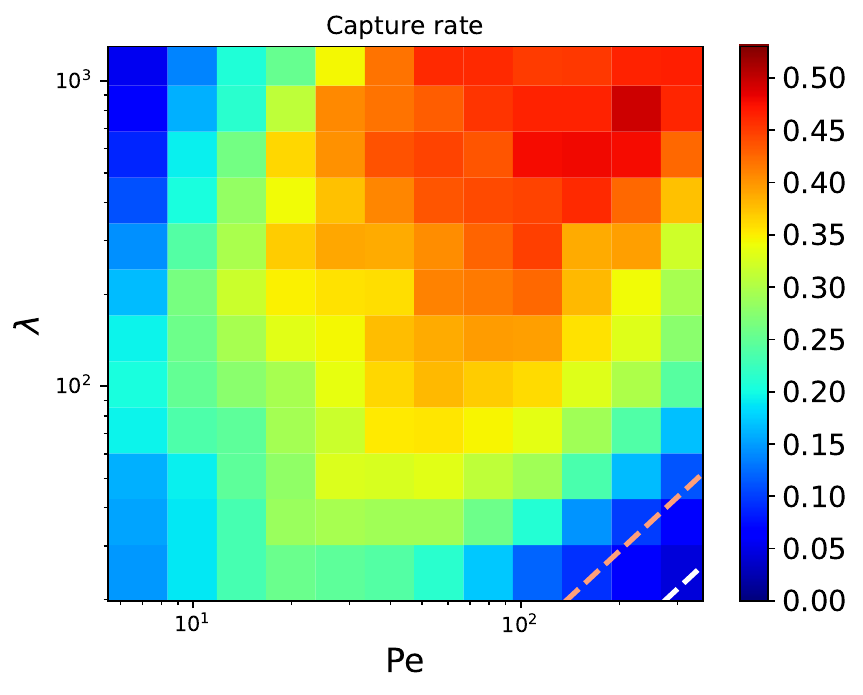}
    \caption{Capture rate histogram depending on adimensional numbers Pe and $\lambda$, after a uniform magnetic field is turned on for a duration $t_B \propto \sigma/v_0$. Each data point is averaged over 20 simulations. The white line represents the stability criterion (\ref{eq:stability}) for $n=1$ and the orange line for $n\to \infty$.}
    \label{fig:phasediagram_fieldduration}
\end{figure}

Figure \ref{fig:phasediagram_fieldduration} is an additional capture rate histogram, but the duration of the magnetic field is scaled to the activity ($t_B \propto 1/\text{Pe}$). In this case, the specific regions of low and high capture rate are seen to have been altered compared to figure \ref{fig:phasediagram_field}. 

For lower values of activity (Pe$< 50$), capture rate increases with Pe. Several phenomena could explain this, such as the increased relative importance of noise in low activity systems or the inability of chains to push passive particles. However, no definite conclusion can be made because another phenomenon is predominant: the duration of the relaxation phase after the magnetic field is turned off is not scaled to the activity. This results in capture rates evaluated before structures become completely inert and unable to capture passive colloids.

For higher diffusive activity (Pe$ > 50$) and lower dipolar interactions ($\lambda < 300$), the capture rate depends on both $\lambda$ and Pe, with dipolar interactions improving the capture while activity hinders it. This suggests that the capture depends on a competition between both effects.

\begin{figure}
    \centering
    \includegraphics[width=0.8\linewidth]{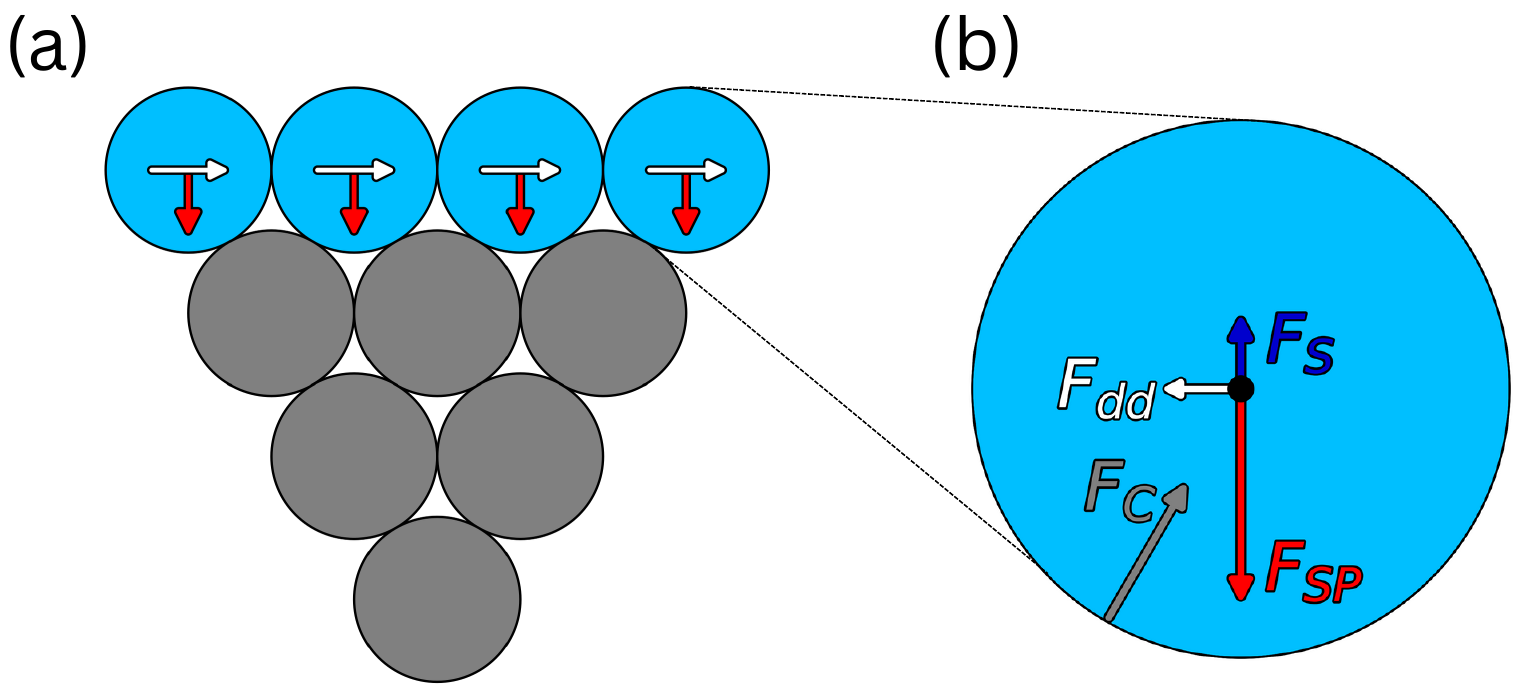}
    \caption{Simplest structure sweeping $n=3$ layers of passive colloids while the magnetic field is on. The structure is stable if, for the active dipole on the edge of the chain, the force $F_{dd} \propto\lambda$ is greater than the force due to the contact with the passive colloid $F_c \propto \frac{n}{n+2}\text{Pe}$.}
    \label{fig:structureforces}
\end{figure}

\section{Discussion}

The competition between activity and dipolar effects may happen in either of the two phases of the protocol, but is more easily understandable when the magnetic field is turned on. An argument based on the stability of the most simple structure that sweeps passive particles can be made to evaluate this competition, as illustrated in figure \ref{fig:structureforces}. This structure is composed by a chain of $n+1$ active dipoles pushing $n(n+1)/2$ passive colloids organized on $n$ layers. For such a structure to not decay into its equivalent of $n-1$ layers of passive colloids, we get the criterion
\begin{align}
    \lambda > \frac{\sqrt{3}}{12} \frac{n}{n+2} \text{Pe} \label{eq:stability}
\end{align}
which gives a linear relation between $\lambda$ and Pe. Two cases are of particular interest. If $n=1$, this relation gives a criterion for the stability of a chain pushing a single passive colloid. It becomes $\lambda < \frac{\sqrt{3}}{36} \text{Pe}$ and indicates that no passive colloid can be captured through this sweeping mechanism. If $n \to \infty$, the relation indicates wether the dipolar interactions are strong enough to hold a structure of a pyramid of any number of layers. This becomes $\lambda > \frac{\sqrt{3}}{12} \text{Pe}$ and indicates that the structures will always stay stable during these sweeping phases. The relation (\ref{eq:stability}) gives an idea on how the phenomena are driven through a competition between $\lambda$ and Pe. These two cases are highlighted in figure \ref{fig:phasediagram_fieldduration}. While explaining under which condition significant capture becomes possible, this reasoning only gives an intuition on how both effects contribute to the actual capture, and not an explanation on how effective the capture is depending on $\text{Pe}$ and $\lambda$. Additional analysis of the growth of active chains, how passive colloids are swept and how active chains fuse with each other must be done to completely understand the competition between activity and dipolar interactions that leads to the capture.

\begin{figure}
    \centering
    \includegraphics[width=0.9\linewidth]{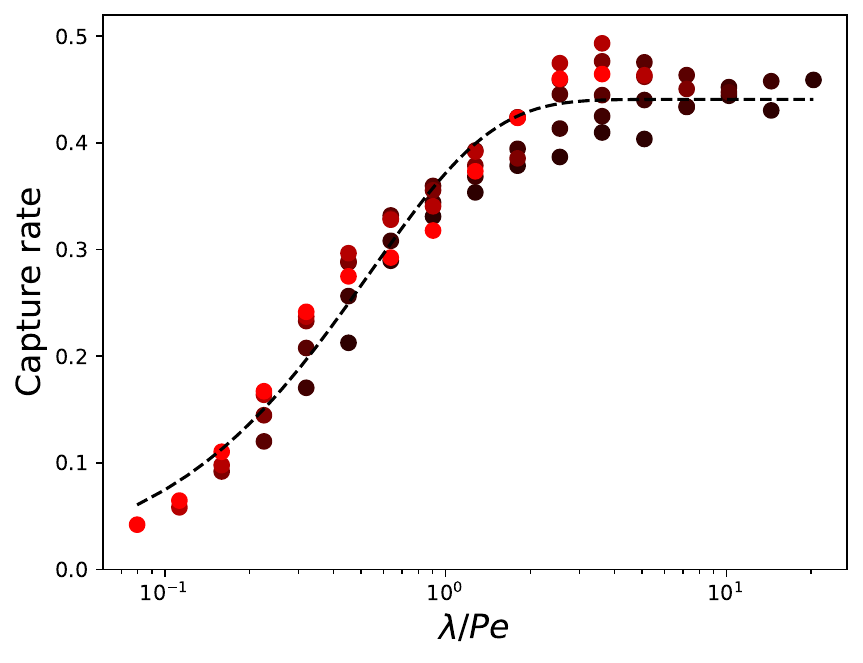}
    \caption{Capture rate of passive colloids depending on $\lambda$/Pe in log-scale for Pe $> 50$, and fit to the saturating function \text{$C  (\frac{\lambda}{\text{Pe}}) = C_{s} \left(1 - \exp(- r_s \frac{\lambda}{\text{Pe}})\right)$} with $C_{s}=0.441$ and $r_s=1.853$. The brighter red a point is, the higher the activity Pe of the system.}
    \label{fig:capture_lbdape}
\end{figure}

The criterion (\ref{eq:stability}), while not capturing all the information about how the trapping happens, can be used to infer that $\lambda/\text{Pe}$ is the leading parameter that determines the results, instead of $\lambda$ and Pe separately. Studying the capture depending on $\lambda/\text{Pe}$, represented in figure \ref{fig:capture_lbdape}, makes it clear that this is the case for Pe $>50$. Capture rate for the same values of $\lambda/\text{Pe}$ are not different enough compared to the variance of the results to draw a clear conclusion that it has a significant effect. This is especially the case since lower capture rates are usually associated to lower values of activity, which may still be impacted by the limitation of relaxation time, although minimal.

Analyzing the capture rate depending on $\lambda/\text{Pe}$ also makes it visually clear that the capture process saturates. By fitting to a saturating exponential function of the form \text{$C_{s} \left(1 - \exp(- r_s \frac{\lambda}{\text{Pe}})\right)$}, we may obtain values for the maximum capture $C_{s}$ and the characteristic rate $r_s$. Both parameters are constant for Pe and $\lambda$, but may depend on other parameters of the system which were kept fixed in this study.

\section{Conclusion}

Finding ways to control passive particles without any other characteristics than steric repulsion may be challenging, especially at low concentrations \cite{stenhammar_activity-induced_2015,wittkowski_nonequilibrium_2017}. External agents, through their activity and other properties, are one way of achieving this objective \cite{gao_seawater-driven_2013,jurado-sanchez_self-propelled_2015}. As seen in this case, complex active dipolar particles may act as sweeping agents for the passive colloids, but the sweeping and capture are done efficiently only if the dipoles are perpendicular to the self-propulsion direction.

While active perpendicular dipoles placed randomly are able to create dense phase of passive colloids through this sweeping effect, the efficiency can be further improved thanks to protocols of external magnetic fields applied globally to the system. A simple temporary activation of external magnetic field is enough to greatly increase the sweeping of passive colloids and their capture in dense phases. 

To further improve this capture, an analysis about how the active and dipolar properties of the active dipoles influence the capture allows us to understand that dipolar interactions favor this effect, while the activity disfavors it, reaching a saturating plateau after a critical value of $\lambda/\text{Pe}$. However, lowering the activity also increases the time scale for which these effects are observed, preventing us to conclude on the actual phenomena happening for $\text{Pe}<50$. This still allows us to find an optimal region of parameters $\lambda$ and $\text{Pe}$ that maximizes the capture.


\vskip 0.2 cm
\section*{Acknowledgments} 

This work is financially supported by the University of Li\`ege through the CESAM Research Unit. 
 

\bibliography{Active_Dipoles}






 \end{document}